\documentstyle[preprint,prl,aps,epsf]{revtex}

\begin{document}
\draft

\title{Old and new results on multicritical points}
\author{Amnon Aharony}
\address{School of Physics and Astronomy, Raymond and Beverly Sackler
Faculty of Exact Sciences, \\ Tel Aviv University, Tel Aviv 69978,
Israel\\ E-mail: aharony@post.tau.ac.il}

\date{\today}
\maketitle
\begin{abstract}
Thirty years after the Liu-Fisher paper on the {\bf bicritical}
and {\bf tetracritical} points in quantum lattice gases, these
multicritical points continue to appear in a variety of new
physical contexts. This paper reviews some recent multicritical
phase diagrams, which involve  e. g. high-$T_c$ superconductivity
and various magnetic phases which may (or may not) coexist with
it. One recent example concerns the SO(5) theory, which combines
the 3-component antiferromagnetic and the 2-component
superconducting order parameters. There, the competition between
the isotropic, biconical and decoupled fixed points yields
bicritical or tetracritical points. Recalling old results on the
subject, it is shown that the {\bf decoupled} fixed point is
stable, implying a {\bf tetracritical} point, contrary to recent
claims, which are critically discussed.  Other examples,
concerning e. g. the superconducting versus charge  and spin
density wave phases are also discussed briefly.
 In all cases, extensions of old results
 can be used to correct new claims.

\end{abstract}
\pacs{KEY WORDS: Multicritical points; bicritical point;
tetracritical point; renormalization group; decoupled fixed
point.}

\section{Introduction}

In addition to celebrating Michael Fisher's 70'th birthday, this
year we also celebrate thirty years to the famous Wilson-Fisher
paper\cite{wf} on the $\epsilon-$expansion. That paper appeared a
few months before I arrived as a post-doc in Fisher's group at
Cornell, and shaped much of my scientific activity in the
following few years. The present paper is dedicated to Michael
Fisher, in recognition of his many contributions to statistical
physics, in gratitude for the many things which I learned from him
in those years and in the 30 years that followed, and in
appreciation for his personal guidance and friendship.

The Wilson-Fisher paper started three decades of activity, in
which the $\epsilon-$expansion was used for many types of
interactions, and for many types of order parameters. In this
paper I concentrate on one special class of these studies,
involving {\bf bicritical} and {\bf tetracritical} points, which
arise when a varying anisotropy causes a crossover from the
critical behavior of an isotropic $n-$component order parameter to
those of order parameters with less components, and hence with
lower symmetries\cite{pfeuty,wegner}. Fisher himself started the
modern theoretical era in this field in his paper with Liu (also
written thirty years ago)\cite{liu}, which gave a detailed mean
field analysis for the case of the supersolid. He then wrote many
more papers on the subject\cite{fn,nkf,knf,mef,dnf,df}. For the
purposes of the present discussion I would like to emphasize his
papers with Kosterlitz and Nelson, on the bi-- and tetracritical
points in anisotropic antiferromagnetic systems\cite{nkf,knf}.

Bi-- and tetracritical points have been revisited quite often
during the last twenty years, whenever new physical systems
required such studies. Here I give a critical review of some
recent discussions of such multicritical points, in the context of
the materials which exhibit high temperature superconductivity.
While parts of the recent literature require new studies of bi--
and tetracritical points, it turns out that many of the ``new"
questions were already discussed in the seventies. The present
paper aims to bridge between the two relevant communities, relate
some of the ``new" questions to some ``old" answers, and
illuminate some questions which still require further study.

\section{Historical review}

The first detailed mean field analysis of bicritical and
tetracritical points was given by Liu and Fisher \cite{liu}. In
that case, the competing order parameters involve the superfluid
and the crystal, within a quantum lattice gas model. They found
three basic scenarios: In the simplest case, the two ordered
phases meet at a first order transition line, which ends at a {\bf
bicritical} point (where the two critical lines between these
phases and the disordered high temperature phase also meet). At
this point, both order parameters become critical simultaneously.
Alternatively, the two ordered phases are separated by a {\bf
mixed} ``supersolid" phase, bounded by two critical lines which
meet the two disordering critical lines at a {\bf tetracritical}
point. The third scenario, which required special choices of the
parameters, is a mixture of the first two: a ``bubble" of a mixed
phase exists near the tetracritical point, ending at some lower
temperature, turning into a first order transition. Being based on
mean field theory, all the expressions for the phase boundaries
are analytic in the parameters (temperature and pressure), and the
lines reach the multicritical point at finite angles with each
other.

Bi-- and tetracritical points were studied extensively in the
context of the anisotropic antiferromagnet (AAFM) in an external
uniform field \cite{rohrer}. In that case, one observes
longitudinal ordering along the easy axis at low fields, with a
first order spin-flop transition into a phase with transverse
ordering. Kosterlitz, Nelson and Fisher \cite{nkf,knf} (KNF) gave
a detailed renormalization group (RG) analysis of this problem,
with both a uniform and a staggered field, and found a rich
variety of phase diagrams, involving both bi-- and tetracritical
points. Beginning with the two order parameter vectors ${\bf S}_1$
and ${\bf S}_2$, with $n_1$ and $n_2$ components, respectively
(with $n_1=1$ and $n_2=n-1$ for the AAFM problem), they wrote the
Ginzburg-Landau-Wilson Hamiltonian
\begin{eqnarray}
{\cal H} &=&-\int d^dx[\frac{1}{2}(r_1{\bf S}_1^2+r_2{\bf
S}_2^2+(\nabla {\bf S}_1)^2+(\nabla {\bf
S}_2)^2)\nonumber\\
&+&u|{\bf S}_1|^4+v|{\bf S}_2|^4+2w|{\bf S}_1|^2|{\bf S}_2|^2],
\label{H}
\end{eqnarray}
and studied the RG flow of $u,~v$ and $w$ on the critical surface,
to first order in $\epsilon=4-d$, where $d$ is the dimensionality
of space. When both order parameters are critical (i. e. at the
multicritical point), there exist six fixed points in the $u-v-w$
space, and the critical behavior is determined by the stable fixed
point, which is approached under the RG flow from some basin of
attraction. At order $\epsilon$, KNF drew a diagram which
indicated which fixed point is stable for different values of
$n_1$ and $n_2$. As either of these numbers increases, stability
switched from the {\bf isotropic Heisenberg fixed point} (IFP)
(with $u^\ast=v^\ast=w^\ast$, hence with full rotational symmetry
in the full $n=n_1+n_2-$component order parameter space), via the
{\bf biconical fixed point} (BFP), (with non-zero $u^\ast,~v^\ast$
and $w^\ast$, representing some lower symmetry) and then to the
{\bf decoupled fixed point} (DFP), at which $w=0$ and each order
parameter has its own critical behavior, similar to that on the
corresponding critical line. As was already known from related
studies \cite{cubic}, the IFP is stable for $n<n_c=4-2
\epsilon+{\cal O}(\epsilon^2)$. Thus, at $d=3$ and $n=3$ one is
close to the stability boundary between the IFP and the BFP. When
the initial parameters are not in the basin of attraction of the
stable fixed point, the system never has an infinite correlation
length, and therefore the transition has been identified as having
a {\bf fluctuation driven first order} \cite{ma,mukamel}.
Quantitatively, one can calculate the details of this transition
by following the RG flow until all the fluctuations are integrated
over, and then treating the resulting free energy (which is
unstable at quartic order, and thus requires the addition of
higher order terms) using a mean field analysis.

The detailed type of the multicritical point (i. e. bi-- or
tetracritical) is determined by the combination $\Delta=uv-w^2$:
this point is tetracritical when $\Delta>0$, and bicritical when
$\Delta \le 0$. KNF thus concluded that one should expect a
bicritical point for the stable IFP, and a tetracritical point for
the stable BCP. The latter also follows for the stable DFP, when
the two critical lines just cross each other. However, the latter
is not relevant for the AAFM, with $n=3$, and therefore has not
been considered in detail.

The shape of the critical lines as they approach the multicritical
point is determined by {\bf scaling}. If the quadratic anisotropy
has the form $g(n_2{\bf S}_1^2-n_1{\bf S}_2^2)$, then the critical
lines approach the multicritical point tangentially, as
$|T_i(g)-T_c| \sim g^{1/\psi_i}$. For the critical disordering
lines, $\psi_i=\phi_g$, where $\phi_g=\nu \lambda_g$ and
$d-\lambda_g$ is the anomalous scaling dimension of $g$; under the
RG iterations, $g(\ell)=e^{\lambda_g \ell}g(0)$, where $e^\ell$ is
the length rescaling factor \cite{pfeuty,wegner,rg}. However, when
the bicritical point is characterized by the IFP, then the
detailed phase diagram below the bicritical point may depend on
the initial value of the parameter $\Delta$. Although $\Delta$ is
{\bf irrelevant} in the RG sense near the IFP, it has slow
transients which decay as $\Delta \sim e^{\lambda_\Delta \ell}$,
with $\lambda_\Delta<0$. If initially $\Delta(0)>0$, then after a
finite number of iterations $\ell$ one may still have
$\Delta(\ell)>0$, resulting with two critical lines bounding a
mixed phase, {\bf as near a tetracritical point!} However, the
difference between these two lines vanishes with $\Delta(\ell)$,
and therefore the exponents $\psi_i$ describing them contain a
combination of $\phi_g$ and of $\phi_\Delta=\nu \lambda_\Delta$
\cite{bruce}. The two critical lines below $T_c$ thus approach
each other {\bf faster} than those above $T_c$. If $\Delta(0)$ is
already small, then one might mistakenly identify these two lines
with a single first order line, and the tetracritical line with a
bicritical one.

As stated, KNF found that to order $\epsilon$, there is always
{\bf only one stable fixed point}. This fact was placed in a more
general context by Br\'ezin {\it et al.} \cite{brezin}, who proved
this statement for {\bf any} quartic combination of the order
parameter components. In related work, Wallace and Zia
\cite{wallace} showed that to order $\epsilon^3$ (at least for
$n>0$) the RG flow is like that of a particle moving in a
potential, with fixed points interchanging stability as they cross
each other in the parameter space. Indeed, all the existing
analyses of such flows (with the exception of $n=0$, where one of
the two stable fixed points cannot be reached for physical reasons
\cite{rg}) always find at most one stable fixed point, even at
higher orders in $\epsilon$. Detailed examples concern the cubic
case\cite{cubic} and the more general $nm-$component order
parameter case, where one can follow these interchanges between
fixed point stabilities in detail \cite{rg}.

Unlike the stability analysis of most fixed points, which relies
on calculations of the stability exponents $\lambda_i$ within the
$\epsilon$ expansion, or numerically, it was realized quite early
that one can discuss the stability of the DFP quite generally,
using {\bf non-perturbative scaling arguments} \cite{rg}. At the
DFP, the coupling term $w|{\bf S}_1|^2|{\bf S}_2|^2$ scales like
the product of two energy-like operators, having the dimensions
$(1-\alpha_{n_i})/\nu_{n_i}$, where $\alpha_{n_i}$ and $\nu_{n_i}$
are the specific heat and correlation length exponents of each
order parameter separately. Thus, the combined operator has the
dimension $d-\lambda_D$, where
\begin{equation}
\lambda_D=\frac{1}{2}\big(\frac{\alpha_{n_1}}{\nu_{n_1}}+\frac{\alpha_{n_2}}{\nu_{n_2}}\big)
\label{stability}
\end{equation}
is the scaling exponent which
determines the RG flow of the coefficient of this term, $w$, near
the DFP.

 Indeed, such arguments gave the
RG basis for the Harris criterion for quenched random systems
(where the parameter which measures the randomness in the coupling
constants scales with $\lambda=\alpha/\nu$) \cite{rg}, and led to
the prediction of a tetracritical point for a quenched random
alloy of systems with competing spin anisotropies \cite{fishman}.
For the two order parameter problem discussed by KNF, one
concludes that in $d=3$ the boundary of stability between the BFP
and the DFP occurs in fact at much lower values of $n_1$ and $n_2$
than those expected from the order$-\epsilon$ estimates.

\section{High temperature superconductors and the story of SO(5)}

The cuprate--based materials exhibit very rich phase diagrams, and
it is generally believed that a good theory should not only
explain the high--temperature superconductivity, but also explain
the other phases which exist near or simultaneously with the
superconducting one. In this connection, it was emphasized already
in 1988 that doping introduces quenched randomness, with a
potential magnetic spin glass phase \cite{ABCKS}. In fact, this
spin glass phase exhibits interesting scaling of the equation of
state \cite{sgexp}, with interesting crossover to a lower symmetry
of the order parameter due to the magnetic field \cite{ravi}. The
concentration--temperature phase diagram presented in Ref.
\cite{ABCKS}, containing many of the interesting phases which
arise in these exciting materials, was later reproduced by Michael
Fisher \cite{gibbs}, to demonstrate possible deviations from the
general Gibbs rules in quenched random systems (Note the improved
graphics introduced by Fisher in this reproduction!)

In its simplest version, the phase diagram of these materials
contains only antiferromagnetic (AFM) and $d-$wave superconducting
(SC) order. In 1997, Zhang\cite{zhang} constructed an SO(5)
theory, which aimed to unify the 3-component AFM order parameter
and the 2-component complex SC order parameter into a combined
5-component theory. As the concentration of electronic holes
increases from half filling (on the copper ions), one expects a
transition from the AFM phase into the SC phase. Zhang argued that
this transition is first order, ending at a bicritical point. This
point is also where the two critical lines, with the critical
behavior of the 3-component AFM and the 2-component SC ordered
phases, meet, ending up with the critical behavior of the higher
symmetry SO(5) group.

In its simple classical version, this SO(5) model maps onto the
model discussed in the previous section, with ${\bf S}_1$ and
${\bf S}_2$ representing the 3-component and 2-component AFM and
SC order parameters, respectively. Indeed, following Zhang's paper
there appeared several papers which repeated some of the RG
analysis reviewed above \cite{BL,MN}, with similar results. In
particular, these references followed the order$-\epsilon$
analysis of KNF, and concluded that for $d=3,~n_1=3$ and $n_2=2$
one has a {\bf tetracritical} point, governed by the {\bf
biconical} fixed point. However, since for $n=5$ the BFP and the
IFP may be close to each other, it has been suggested that one
might actually observe a bicritical point, with exponents
dominated by the IFP. Measurement of such exponents was even
presented as ``an experimental measurement of the number 5 of the
SO(5) theory"! \cite{zhang1} However, even in such a scenario,
Ref. \cite{MN} incorrectly stated that the all the four phase
boundary exponents $\psi_i$ are the same, equal to $\phi_g$ (in
contrast to Ref. \cite{bruce}).

Following this background, Hu~\cite{hu} used Monte Carlo (MC)
simulations on an SO(5) rotator model, and concluded that the
multicritical point which characterizes the simultaneous ordering
of the SO(3) AFM 3-component and of the U(1) SC 2-component order
parameters, ${\bf S}_1$ and ${\bf S}_2$, has the critical behavior
of the {\it isotropic} 5-component rotator model. This seems to
contradict the RG in $d=4-\epsilon$ dimensions, which states that
(a) to a high order in $\epsilon$, the isotropic SO($n$) fixed
point (IFP) is {\it unstable} for $n>n_c$, with $n_c<4$ \cite{rg},
and (b) to order $\epsilon$, this multicritical point is described
by the anisotropic biconical fixed point \cite{BL,MN,knf}.

These MC results by Hu (as well as the statements in many of the
SO(5) papers in the literature) suffer from several problems.
First, one might question the relevance of this discussion to
high-$T_c$ superconductivity (where one should also include
fluctuations in the {\bf electromagnetic gauge field}~\cite{ma}).
Second, these papers ignore the {\bf quenched randomness}, which
is intrinsic for all of the doped cuprates (even if some
electronic properties may be viewed as dominated by extended wave
functions). Here we ignore these two points, and concentrate on
the third issue: as reviewed in the previous section,
 at $d=3$ the multicritical point {\it must} be
tetracritical, being characterized by the {\it decoupled} fixed
point (DFP). Returning to Eq. (\ref{stability}), we can now use
the known negative values of $\alpha_2$ and $\alpha_3$ at
$d=3$~\cite{legui}, to find that $\lambda_D\cong-0.087<0$, and the
DFP is stable, in contrast to the order-$\epsilon$ extrapolation
to $\epsilon=1$ \cite{BL,MN,knf}. Thus, asymptotically the free
energy breaks into a sum of the two free energies, ${\bf S}_1$ and
${\bf S}_2$ exhibit the Heisenberg ($n=3$) and XY ($n=2$) critical
exponents and the two critical lines cross each other at finite
angles, with the crossover exponent $\phi=1$ \cite{comment}. The
latter statement is only asymptotic; after a finite number of RG
iterations one still has a finite $w(\ell)$, yielding corrections
to the phase boundaries which approach the asymptotic lines
tangentially. Accurate experiments in the {\it asymptotic} regime
thus carry no information on the SO(5) theory. However, they may
yield some information on the transient non-asymptotic behavior
near the {\it initial} Hamiltonian.

Ref. \onlinecite{hu} used a {\it discrete} spin model, with $|{\bf
S}_1|^2 +|{\bf S}_2|^2=1$. This is believed to be in the same
universality class as a Ginzburg-Landau-Wilson (GLW) theory, with
the quartic term $U(|{\bf S}_1|^2+|{\bf S}_2|^2)^2$ (where
initially $U\longrightarrow\infty$)~\cite{WK}. Ref.
\onlinecite{hu} then added a coupling $W|{\bf S}_1|^2|{\bf
S}_2|^2$. Quantum fluctuations~\cite{hanke} and RG
iterations~\cite{rg} then also generate a term $V(|{\bf
S}_1|^4-|{\bf S}_2|^4)$. Clearly, $u=U+V,~v=U-V$ and $w=U+W/2$.
Again, there exist six fixed points in the $U-V-W$ parameter
space, of which {\it only one} should be
stable~\cite{rg,brezin,wallace}. For a continuous transition, the
above argument implies an RG flow away from the vicinity of the
unstable IFP, at $V=W=0$, to the DFP, where $2U+W=0$. This flow
may be slow, since the related exponents $\lambda^V_I$ and
$\lambda^W_I$ are small: the asymptotic DFP behavior can be
observed only if $W X^{\lambda^W_I}$ becomes comparable to $U$,
which is large. Here, $X=\min(L,\xi)$, with $L$ the sample size
and $\xi \sim(T-T_c)^{-\nu}$ the correlation length ($T_c$ is the
temperature at the multicritical point).
 Therefore, one might need to
go very close to the predicted tetracritical point, and to much
largewr samples, in order to observe the correct critical
behavior. The simulations of Ref. \onlinecite{hu}, which begin
close to the ITP ($U\gg V,~W$) and use relatively small $L$,
apparently stay in the {\it transient} regime which exhibits the
isotropic exponents. To observe the true asymptotic decoupled
behavior, one should start with a more general model, allowing
different interactions for ${\bf S}_1$ and for ${\bf S}_2$, relax
the strong constraint $|{\bf S}_1|^2 +|{\bf S}_2|^2=1$, and use
much larger $X$. The latter is also needed due to the small value
of $\lambda_D$. These requirements may be impossible for realistic
MC simulations. In fact, Hu recently generalized his MC
simulations, and used {\bf finite} values of $U$ \cite{hureply}.
However, his initial parameters obeyed $W<4U$, which may still be
much too close to the IFP. In these additional simulations, Hu
still finds a bicritical phase diagram, with critical exponents
which seem close to those of the IFP, thus contradicting the
theoretical asymptotic expectation of a tertacritical point
associated with the DFP.

There are three possible ways to explain this discrepancy:
\begin{itemize}
\item
The crossover due to the RG flow from the initial vicinity of the
IFP to the asymptotic DFP could be too slow, requiring much larger
values of $X$ than practical in the simulation. As $X$ increases,
I would expect signals of approaching the DFP. An example of such
a signal would be the appearance of a ``bubble" of the mixed phase
near the multicritical point. Since this bubble may be narrow (and
short), it could easily by identified as a single first order
transition line. At low temperatures, the bubble could close back
into the first order line, e. g. due to higher order terms in the
Ginzburg-Landau expansion (as happened e. g. in the Liu-Fisher
phase diagram).

\item
If the initial Hamiltonian were out of the basin of attraction of
the DFP, then one should observe first order transitions from the
disordered phase into the ordered phases \cite{MN}. Again, the
discontinuity on these transitions may be too small for the
available values of $X$.

\item
Finally, there could be {\bf two} stable fixed points. As stated
several times above, I find this scenario most unlikely. In
particular, it is well established that the IFP is {\bf unstable}
for $n>4$. However, if indeed this scenario turns out to be true,
then this case would represent a mini-revolution in our thinking
of  RG flows in such systems. It would be nice to have
generalizations of the Br\'ezin {\it et al.} and of the Wallace
and Zia arguments to all orders in $\epsilon$, and specifically
for $d=3$.

\end{itemize}

\section{Other examples}

In addition to the simple AFM ordering, there have been many
recent scattering experiments which exhibit some kind of (static
or dynamic) {\bf incommensurate peaks} \cite{expt}. These peaks,
which may correspond to density and/or spin density wave ordering,
usually arise at doping concentrations above those of the AFM
phase, and often coexist with the SC phase. The general theory
discussed above can thus be transferred to this new competition.
In the simplest case, ${\bf S}_1$ would represent the {\bf spin
density wave} (SDW) order parameter, and ${\bf S}_2$ would
continue to represent the SC ordering. Indeed, Kivelson {\it et
al.} \cite{kivelson} generated a variety of
temperature-concentration phase diagrams, taking account of the
fact that the concentration $x$ is related to the chemical
potential $\mu$ which appears in the Ginzburg-Landau Hamiltonian
via a Legendre transform. At the moment, there exists no detailed
RG analysis of this case, which should be a generalization of the
Fisher-Nelson \cite{fn} treatment of the AAFM at fixed
magnetization. Apart from taking note of the ``old" literature,
such an analysis should also be careful in counting the components
of the SDW order parameter. For an incommensurate wave vector,
this number could be significantly larger than three
\cite{mukamel}, and the RG may not have a stable fixed point at
all, implying a fluctuation driven first order transition.

One theoretical scenario for the SDW ordering concerns {\bf
stripes}, which involve {\bf charge density waves}
\cite{kivelson}. This leads to a three-fold competition, between
SDW, CDW and SC \cite{zachar}. Since the wave vector of the CDW is
equal to twice that of the SDW, this yields terms which are linear
in the CDW order parameter and bilinear in the SDW one, possibly
leading to first order transitions into the CDW phase
\cite{zachar}. Again, both the CDW and the SDW can have a large
number of components, turning the RG treatment (not yet done)
complicated but interesting.

Finally, I mention another class of phase diagrams, involving
superconductivity in the bismuthates \cite{AA}. These systems
exhibit both CDW and SC ordering, and their
temperature-concentration phase diagrams have drawn much interest
even before the discovery of high temperature superconductivity
\cite{rmp}. It turns out that both types of order can follow from
a negative-$U$ Hubbard model, which can then be mapped onto an
anisotropic Ising-Heisenberg spin model. A mean field analysis of
this model\cite{AA} yields phase diagrams which are similar to
those found by Fisher and Nelson\cite{fn}, with their
magnetization replaced by the concentration. The resulting
coexistence region was ignored in earlier analyses\cite{rmp}. In
addition, the quenched randomness generates effective {\bf random
fields}, which couple to the CDW order parameters and cause a
breakdown of that phase into finite domains, as apparently
observed experimentally. It would be interesting to search for
similar effects in the cuprates. It would also be interesting to
have a comprehensive study of the role played by quenched
randomness in these interesting systems.

\section{Conclusions}

\begin{itemize}
\item New materials bring about new phase diagrams, with competing
types of order and with a variety of multicritical points.
Cuprates and bismuthates are good examples of such rich varieties.
\item
Many details of these phase diagrams are often available from the
early days of the RG research. It would help to bridge between the
SC and the RG communities.
\item
In the context of SO(5), it would help to have more accurate
experiments, as well as more MC simulations, in regimes which
might be better suited for reaching the asymptotic correct
behavior. In parallel, it might be of interest to find ways to
investigate the relative stability of the competing fixed points
{\bf at} $d=3$.
\item
After 30 years of RG studies, there are still new problems which
require new RG treatments. It is appropriate to celebrate Fisher's
70th birthday recalling his ``old" contributions, which opened the
way to much of this ``new" activity.

\end{itemize}

\noindent {\bf Acknowledgements}

 This work was supported by the U. S. -- Israel Binational Science Foundation (BSF)
 and by the German-Israeli Foundation (GIF).

 \end{document}